\documentclass[usegraphicx]{mn2e}
\usepackage{subfigure}
\usepackage{epsfig}
\usepackage{amsmath}
\usepackage{amssymb}
\usepackage{graphicx}
\usepackage{times,longtable,color}
\usepackage[T1]{fontenc}
\usepackage{aecompl}

\newcommand {\apgt} {\ {\raise-.5ex\hbox{$\buildrel>\over\sim$}}\ }
\newcommand {\aplt} {\ {\raise-.5ex\hbox{$\buildrel<\over\sim$}}\ } 
\newcommand {\degree}{$^{\circ}$}

\title[The view of AGN-host alignment via reflection spectroscopy]
{The view of AGN-host alignment via reflection spectroscopy}

\author[M.Middleton et al.]
{Matthew J. Middleton$^{1}$, Michael L. Parker$^{1}$, Christopher S. Reynolds$^{2, 3}$, \newauthor Andrew C. Fabian$^{1}$ and Anne M. Lohfink$^{1}$\\
\\
1. Institute of Astronomy, Madingley Road, Cambridge CB3 OHA\\
2. Dept. of Astronomy, University of Maryland, College Park, MD20742, USA\\ 
3. Joint Space-Science Institute (JSI), College Park, MD 20742- 2421, USA
}

\pagerange{\pageref{firstpage}--\pageref{lastpage}} \pubyear{2011}
\long\def\symbolfootnote[#1]#2{\begingroup\def\thefootnote{\fnsymbol{footnote}}\footnote[#1]{#2}\endgroup} 
\def\ga{\mathrel{\hbox{\rlap{\hbox{\lower4pt\hbox{$\sim$}}}{\raise2pt\hbox{$>$}}
}}}

\begin{document}

\topmargin = -0.5cm

\maketitle

\label{firstpage}
\begin{abstract}
The fuelling of active galactic nuclei (AGN) - via material propagated through the galactic disc or via minor mergers - is expected to leave an imprint on the alignment of the sub-pc disc relative to the host galaxy's stellar disc. Determining the inclination of the inner disc usually relies on the launching angle of the jet; here instead we use the inclination derived from reflection fits to a sample of AGN. We determine the distorting effect of unmodeled Fe XXV/XXVI features and, via extensive simulations, determine the difference in disc inclination resulting from the use of {\sc relxill} compared to {\sc reflionx}. We compare inner disc inclinations to those for the host galaxy stellar disc derived from the Hubble formula and, via Monte-Carlo simulations, find a strong lack of a correlation (at $\gg$ 5$\sigma$) implying either widespread feeding via mergers {\bf if} we assume the sample to be homogeneous, or that radiative disc warps are distorting our view of the emission. However, we find that by removing a small ($\sim$1/5) subset of AGN, the remaining sample is consistent with random sampling of an underlying 1:1 correlation (at the 3$\sigma$ level). A heterogenous sample would likely imply that our view is not dominated by radiative disc warps but instead by different feeding mechanisms with the majority consistent with coplanar accretion (although this may be the result of selection bias), whilst a smaller but not insignificant fraction may have been fuelled by minor mergers in the recent history of the host galaxy.

\end{abstract}

\section{Introduction}

Directly observing the inner sub-pc regions of active galactic nuclei (AGN) requires resolving powers beyond the abilities of instrumentation for the foreseeable future; as a result, it is unclear how material is fed into the accretion disc and onto the supermassive black hole (SMBH). It is possible that material is funnelled through the kpc-scale galactic stellar disc (with angular momentum lost via dissipation) into the sub-pc AGN disc or via a merger of a dwarf satellite galaxy, with the gas falling into the inner sub-pc regions, forming a disc and once again feeding the SMBH.  A possible means to distinguish between these two methods - and in doing so determine the mechanism responsible for recent AGN activity - is to search for a general alignment or misalignment of the host galaxy with the inner AGN disc (see Hopkins et al. 2012). We expect the latter to align with the SMBH within the warp radius induced by the Bardeen-Petterson (BP) effect (Bardeen \& Petterson 1975; Pringle 1992; Papaloizou \& Lin 1995) which, for typical SMBH masses, Eddington ratios (and for reasonable values of the viscosity) will be of order 100-1000s R$_{\rm g}$ (Rees 1978; Scheuer \& Feiler 1996; King 2005). 

Our view of alignment or misalignment results from the combination of the relative angular momentum of the infalling material, the timescales for alignment/misalignment (which depends on the amount of angular momentum being transferred: King et al. 2005) and whether the SMBH angular momentum vector ($\bf {\hat{J}_{BH}}$) is aligned or misaligned with that of the host galaxy ($\bf{\hat{J}_{gal}}$) at the time we see it accreting. In the simplest case, observing systematic {\it alignment} of host galaxy and AGN inner disc could be a result of co-planar fuelling via the host stellar disc onto an aligned SMBH - by which we mean that the SMBH's angular momentum vector is parallel to that of the host galaxy ($\bf{\hat{J}_{BH}.\hat{J}_{gal}}$ = $\bf{\mid\hat{J}_{BH}\mid\mid\hat{J}_{gal}\mid}$). Alternatively, apparent alignment could result from a recent merger but with the gas falling onto a SMBH which is misaligned with the host galaxy ($\bf{\hat{J}_{BH}.\hat{J}_{gal}} < \bf{\mid\hat{J}_{BH}\mid\mid\hat{J}_{gal}\mid}$) and where the accretion flow has not yet had time to align with the black hole (i.e. the time required for the BP effect to lead to a flow aligned with $\bf {\hat{J}_{BH}}$) and so appears by chance to be aligned with the host galaxy. {\it Misalignment} of AGN inner disc and host galaxy on the other hand could result from co-planar accretion through the stellar disc onto a SMBH which is itself misaligned with the host galaxy (and which would have required mergers in the past with both gas and the merging SMBH affecting the alignment, e.g. Gerosa et al. 2015). Alternatively, a recent merger could lead to gas falling onto the SMBH with misaligned angular momentum relative to an aligned (with the host) SMBH; if the inflow has not yet had time to align with the SMBH via the BP effect then the inner disc would appear misaligned. Finally, infalling gas during a merger could fall onto a SMBH which is misaligned with the host galaxy and enough time has passed for the BP effect to align the inflow with the SMBH (and therefore be misaligned relative to the host galaxy). It is worth noting that a general requirement for misalignment (as a result of the feeding mechanism) would therefore appear to be mergers driving the recent AGN activity or in the past to misalign the SMBH.

AGN-host misalignment may also result from radiative warping  (Petterson 1977; Iping \& Petterson 1990; Pringle 1996, 1997) where isotropic radiation illuminates the inner disc in a non-uniform manner and results in a non-uniform back-pressure which induces a torque, changes the angular momentum and induces a warp. Providing the warping is slow compared to the black-hole alignment timescale, this is expected to occur at 10$^{5-6}$~R$_{\rm g}$ in AGN although is dependent on the impact of winds on the back-pressure (see Schandl \& Meyer 1994; Schandl 1996) and the mode by which the warp propagates (e.g. Papaloiozou \& Lin 1995). Distinguishing between misalignment induced via radiative warping or as a result of a misaligned SMBH or feeding via minor mergers requires detailed optical studies to search for the signature of disturbed motion on larger scales (e.g. Raimundo et al. 2013; Davies et al. 2014) or for filaments associated with mergers  (e.g. Fischer et al. 2015). In addition, where the structure and behaviour of the accretion flow does not differ greatly between sources, radiative warping is expected to lead to {\it ubiquitous} misalignment. As a result, searches for systematic alignment or misalignment in AGN should provide an opportunity to better isolate the processes occurring in the inner regions, with widespread AGN-host alignment indicating that radiative warping is unimportant and the AGN is likely fed via the host galaxy, whereas misalignment could, in principle, result from either feeding mechanism but requiring radiative warping or a misaligned SMBH if via co-planar accretion (with the feeding mechanism then distinguished by optical studies).

\begin{table*}
\begin{center}
\begin{minipage}{100mm}
\caption{Table of AGN sample parameters}
\begin{tabular}{l l l l }
\hline
\hline
Object	&	$i_\textrm{ref}$	&	Reference	&	$i_\textrm{opt}$	 	 \\ 
\hline															
{\bf 1H 0419-577}	&	$51^{+4}_{-6}$	&	Walton et al. (2013)	& 11 $\pm$ 8	 $^{a}$\\ 
{\bf 1H 0707-495} $^\dagger$	&		$56 \pm 1 ^{+14}_{-6}$	&	Fabian et al. (2009)	&	29 $\pm$ 8	\\ 
{\bf 3C 120}	&			$5^{+4}_{-5}$	&	Lohfink et al. (2013)	 	&	65 $\pm$ 7	 \\ 
Ark 120	&			$54^{+6}_{-5}$	&	Walton et al. (2013)		&	52 $\pm$ 12	\\ 
Ark 564 $^\dagger$  	&			$64^{+1 / +6}_{-11}$	&	Walton et al. (2013)	&	47 $\pm$ 5		\\ 
ESO 362-G18	&			$53\pm5$	&	Agis-Gonzalez et al. (2014)	&	71 $\pm$ 6	\\ 
Fairall 9	&			$48^{+6}_{-2}$	&	Lohfink et al. (2012)		&	64 $\pm$ 8	\\ 
IRAS 00521-7054 $^\dagger$ 	&		$37^{+4/ +13}_{-4 /-7}$	&	Tan et al. (2012) 	&	52 $\pm$ 8	\\ 
{\bf IRAS 13224-3809} $^\dagger$ 	&			$65^{+1 / +5}_{-1}$	&	Chiang et al. (2015)	&	28 $\pm$ 8	\\ 
MCG--06-30-15	&		$33\pm3$	&	Marinucci et al. (2014a)	&	59 $\pm$ 6	 \\ 
Mrk 1018 $^\dagger$ 	&			$45^{+14}_{-10 / -15}$	&	Walton et al. (2013)	&	70 $\pm$ 6	\\ 
Mrk 79	&		$24\pm1$	&	Gallo et al. (2011)		&	37 $\pm$ 6		\\ 
Mrk 110	&			$31^{+4}_{-6}$	&	Walton et al. (2013)		& 47 $\pm$ 8			\\ 
{\bf Mrk 335}	&		$65\pm1$	&	Parker et al. (2014)			&	$<$ 17		\\ 
Mrk 359	&			$47\pm6$	&	Walton et al. (2013)			&	46 $\pm$ 8		\\ 
Mrk 509	&			$<18$	&	Walton et al. (2013)			&	36 $\pm$ 8		\\ 
Mrk 841	&			$45^{+7}_{-5}$	&	Walton et al. (2013)		&	10 $\pm$ 8 $^{b}$	\\ 
NGC 1365	&			$63\pm4$	&	Walton et al. (2014)		&	63 $\pm$ 4	 \\ 
NGC 3227	&			$47_{-2}^{+3}$	&	Patrick et al. (2012)	&	68 $\pm$ 4 	\\ 
NGC 3783	&			$22^{+3}_{-8}$	&	Brenneman et al. (2011)	&	27 $\pm$ 8		\\ 
NGC 4051	&		$13\pm2$	&	Patrick et al. (2012)		&	30 $\pm$ 10		\\ 
NGC 4151	&			$<20$	&	Keck et al. (2015)			&	42 $\pm$ 8		\\ 
NGC 7469 &			$<54$	&	Walton et al. (2013)		&	30 $\pm$ 8		\\
PDS 456	&			$70^{+3}_{-5}$	&	Walton et al. (2013)		&			\\ 
PG 1211+143 $^\dagger$  	&	$28^{+7 / +22}_{-7}$	&	Zoghbi et al. (2015)			&	49 $\pm$ 8 $^{c}$\\ 
RBS 1124	$^\dagger$ &			$66^{+5}_{-15}$	&	Walton et al. (2013)	&			\\ 
Swift J2127.4+5654	&			$49\pm2$	&	Marinucci et al. (2014b)	&			\\ 	
Ton S180	$^\dagger$ &		$60^{+3 / +10}_{-1 / -10}$	&	Walton et al. (2013) &	55 $\pm$ 8		\\
UGC 6728	&		$<55$	&	Walton et al. (2013) &	56 $\pm$ 8		\\ 	

\hline
\hline
\end{tabular}
Notes: Inclinations (and associated literature references) for our AGN sample, with the inner disc inclination ($i_{\rm ref}$) derived from reflection model fitting and the host galaxy stellar disc inclination ($i_{\rm opt}$) mostly sourced from the {\sc HyperLEDA} database (Makarov et al. 2014). Where values are not available for the latter, we obtain measurements for the semi-major and semi-minor axis (and morphological class - where this is not available we assume $t$ = 6) and determine the inclination from the Hubble formula (equation 1). $^{a}$ indicates host galaxy measurements were taken from Asmus et al. (2014), $^{b}$ from Jansen et al. (2000) and $^{c}$ from Brauher et al. (2008). Where the symbol $\dagger$ appears next to a source, narrow absorption or emission was not accounted for in the reflection modelling (usually as a result of the the data not supporting its inclusion). Errors on $i_{\rm ref}$ are typically at 90\% confidence and where two sets are quoted, the larger limit is a rough indication of possible uncertainty arriving from the effect of unmodeled Fe XXV/XXVI in absorption or emission. Errors on $i_{\rm opt}$ were determined from the uncertainty in $logr_{\rm 25}$ and morphological type. Where morphological type was unknown, the inclination assumes $t$ = 6 and we use the mean error from those inclinations where errors on the morphology were available (8\degree). Those sources which we have found to be probable outliers from a 1:1 correlation are indicated in bold.

\end{minipage} 

\end{center}
\end{table*}

The inclination of the host stellar disc can be constrained via the ratio of semi-major to semi-minor axis (for a given morphological type using the classical Hubble formula) or via the tracer of galactic HI emission. The inclination of the inner sub-pc disc is somewhat harder to determine unambiguously; whilst the presence of a molecular torus and the nature of optical emission lines may allow a crude estimate of the inclination of the sub-pc regions via the unified model (Antonucci \& Miller 1985; Antonucci 1993), the orientation of kpc-scale jets has been the traditional means of estimating the inclination under the assumption that jets are launched along the spin axis of the black hole (e.g. Blandford \& Znajek 1977 but see also Natarajan \& Pringle 1998) which is expected to be aligned with the inner disc via the BP effect. The latter technique relies on VLBI measurements of the fastest part of the jet and so the inclination is broadly speaking only an approximation. Further possible means for determining the orientation of the inner flow include the study of those few H$_{2}$O megamasers (e.g. Herrnstein et al. 1999), from the dynamics of the narrow-line region (assuming a bi-conical outflow: Fischer et al. 2013) or from the motion of the broad-line region clouds (e.g. Zhang \& Wu 2002) for an assumed geometry (and as long as velocity dispersion measurements are available), although the reliability of the inclinations via this last method must necessarily depend on the error on the SMBH mass which can be substantial (e.g. Gebhardt et al. 2000). As opposed to the constraints available from the above methods, we can in principle directly {\it measure} the inclination of the inner disc from the signature of reflection of a primary X-ray continuum from neutral to partially-ionised material in the optically thick accretion disc (e.g. Fabian et al. 1989). Such reflection imprints strong Fe K$_{\alpha}$ features in emission, which are broadened and relativistically boosted depending on the inclination of the disc to the observer and the radius of the disc from which the reflected emission originates. The red wing of the Fe K$_{\alpha}$ line is heavily sensitive to the most inner radius from where it is generated and has been widely used as a means to estimate the mass-independent black hole spin in both AGN and black hole binaries (see the reviews of Reynolds 2014; Miller \& Miller 2015; Middleton 2015), whilst the blue wing is most sensitive to the inclination of the inner disc (due to Doppler boosting). However, the inclination can be degenerate with the spin and radial emissivity profile of the reflected emission (although the latter can now be broadly parameterised and understood for a range of geometries: Wilkins \& Fabian 2012; Dauser et al. 2013), whilst the presence of narrow absorption and emission lines (namely Fe XXV/XXVI) can lead to a distortion of the line profile (see the discussion in Lohfink et al. 2012) that has to be carefully modelled (although data quality can sometimes preclude this). 

Here we combine the latest values of the inner disc inclination derived from reflection fits to a sample of (predominantly Seyfert 1) AGN with the inclination of the host stellar disc, and perform a rigorous search for the presence or lack of a correlation. 

\begin{figure*}
\includegraphics[width=7in]{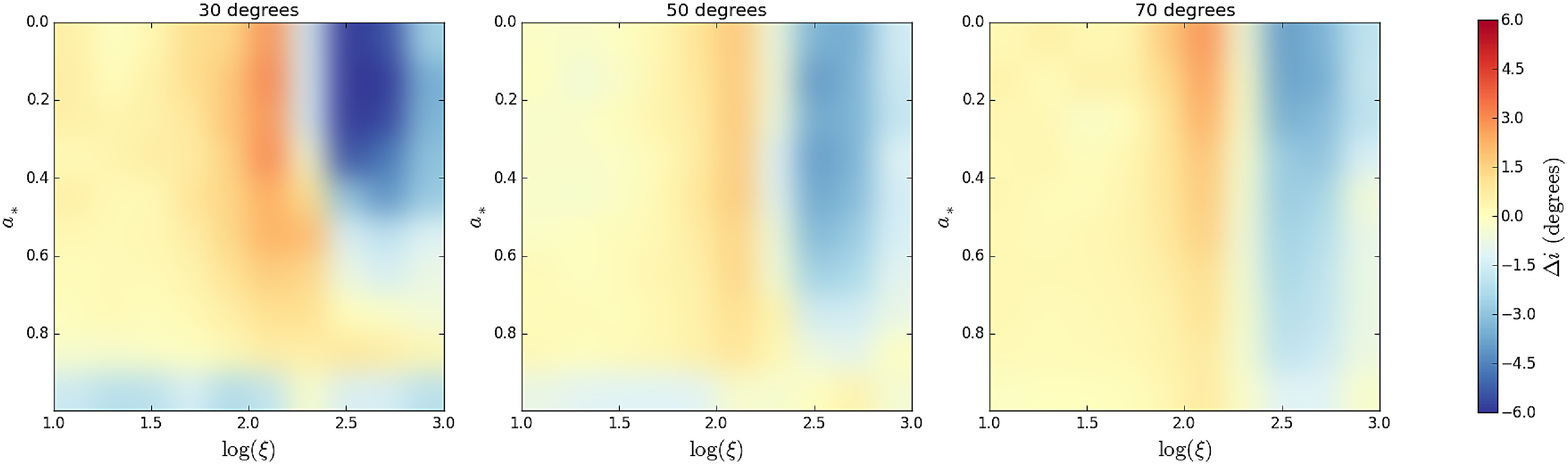}
\caption{Deviation colour map from our simulations using {\sc relxill} as input model and fitting with {\sc reflionx}. We simulate across a range in log$\xi$ (2-3 in 100 even logarithmic steps), for a range of black hole spin values (a$_{*}$ = 0 - 0.98 in 100 steps) and for a range of input inclinations (30, 50 and 70\degree). The maximum deviations are 6.98, 4.10 and 3.97\degree for input inclinations of 30, 50 and 70\degree respectively.
}
\end{figure*}

\begin{figure*}
\includegraphics[width=6in]{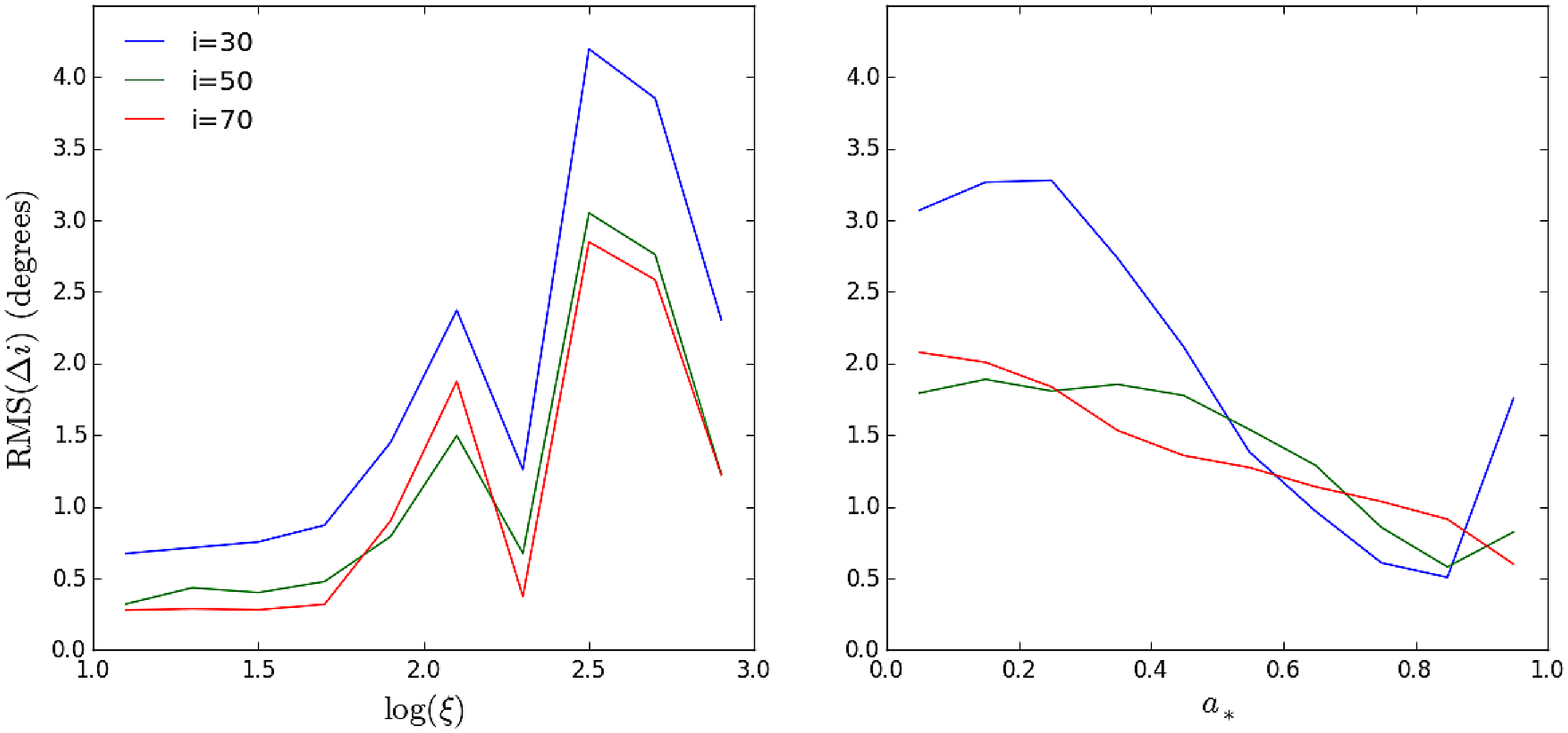}
\caption{As for Figure 1 where the deviation has been collapsed across parameter space to give an indication of mean deviation as a function of ionisation state (left, with values averaged across the range in spin) and black hole spin (right, with values averaged across the range in ionisation state) for the three input inclination values. Averaging across both of these plots gives a mean rms (i.e. the mean deviation for the sample with mean of $\approx$ 0) of 2.22, 1.50 and 1.45\degree for starting inclinations of 30, 50 and 70\degree respectively.
}
\end{figure*}


\section{Sample selection}

We compile values for the inner disc inclination from several recent studies of AGN which utilise current, relativistic reflection models and which apply the criteria discussed in Reynolds (2014), namely that the spectral fits were broad-band and allowed for free metal abundances - this provides a sample of 29 AGN which are given in Table 1. Where duplicates are found, we select inclinations derived from use of the highest quality data (and especially those for which {\it NuSTAR} data was used as this allows the full reflection spectrum to be well modelled and parameters better constrained). 

\subsection{{\sc reflionx} vs {\sc relxill}}

All of the reported inclinations in Table 1 result from the use of either {\sc reflionx} (Ross \& Fabian 2005) or {\sc relxill/xillver} (Dauser et al. 2013) and do not assume a specific geometry (as would be the case with, e.g. {\sc relxill\_lp}: Dauser et al. 2013). Garcia et al. (2013) compare these two reflection models, determining that they are in overall reasonable agreement with less than 30 percent discrepancy at maximum, and which decreases with decreasing ionisation state and softening of the illuminating power-law continuum. Although the impact on the inclination is not explicitly mentioned, the models agree very well around the Fe K edge (important for isolating the strength of the blue shifted Fe K$_{\alpha}$ line) and the equivalent widths of the Fe K$_{\alpha}$ lines have been shown to be similar for low or high ionisations ($2 \lesssim {\rm log}\xi \gtrsim 3$). However, as many AGN in our sample have reported ionisations in the range log$\xi$ 2-3 (see as example, the collated sample of Walton et al. 2013), we proceed to explore the deviation in derived inclinations from using the two reflection models. 

To do this we simulate spectra in {\sc xspec} (using the {\sc fakeit} command) using the {\sc relxill} model for a range of log$\xi$ (2-3 in 100 even logarithmic steps), for a range of black hole spin values (a$_{*}$ = 0 - 0.98 in 100 steps) and for a range of inclinations (30, 50 and 70\degree - the upper limit precludes inclinations likely to intercept the molecular torus). We assume an input power-law index of $\Gamma$=2 to be consistent with the {\it XMM-Newton} and {\it NuSTAR} study of NGC 1365 (Walton et al. 2014) which we take to be broadly representative of our sample (see e.g. Walton et al. 2013). We approximate the emissivity by an illumination pattern following R$^{-7}$ (consistent with an illuminating on-axis source close to the black hole and is the approximate mean value measured for NGC 1365: Walton et al. 2014) breaking to R$^{-3}$ at 6~R$_{\rm g}$ (the ISCO radius for a$_{*}$ =  0) such that increasing the spin (which brings the ISCO inwards) increases the contribution from the regions where light-bending is greatest. We fix the reflected fraction to unity such that for higher spins, our error bounds are conservative (as a higher spin increases the inner disc area, the reflected fraction, and therefore our ability to determine the inclination: Wilkins \& Fabian 2012). We simulate spectral data for 100~ks (a representative length of the best available single-look AGN datasets with {\it XMM-Newton}) across the nominal {\it XMM-Newton}, 0.3-10~keV bandpass (as the majority of the values presented in Table 1 do not benefit from the use of the expanded bandpass {\it NuSTAR} provides) for a representative flux of 10$^{-11}$ erg cm$^{-2}$ s$^{-1}$. The exposure time is not critical but is large enough so that the signal-to-noise ratio is high enough that a global minimum in chi-squared fitting is easily found and is long enough so that the output simulated spectra are close to the input model values (which we confirm by refitting the simulated data with {\sc relxill}). Finally, we fit the simulated data with {\sc reflionx} to determine the uncertainty in the inclination which is plotted for each starting inclination and our ranges in spin and ionisation parameter in Figure 1, with the mean deviations (over the range of spin values and ionisation states) shown in Figure 2. 

Inspection of Figures 1 and 2 indicates that the deviation in measured inclination value is, as expected, a function of input inclination with the largest measured deviations of 6.98, 4.10 and 3.97\degree and mean rms deviations (across {\it both} the range of spin values and ionisation states) of 2.22, 1.50 and 1.45\degree for starting inclinations of 30, 50 and 70\degree respectively. Thus, on average, the errors associated with using {\sc reflionx} rather than {\sc relxill} to determine inclinations are smaller than the measurement errors in Table 1. Ignoring the effects of narrow absorption and emission lines (see the following sub-section), we therefore do not expect substantial differences between the values returned by the two models; as confirmation of this, we note that in the study of Mrk 335, Parker et al. (2014) find consistent inclination values using either model. Finally, we note that our predicted errors are likely to be over-estimates given that we ignore the contribution of the Compton hump which can improve constraints on the inclination (e.g. Garcia et al. 2013) where such higher energy coverage is available.


\begin{figure*}
\includegraphics[width=3in]{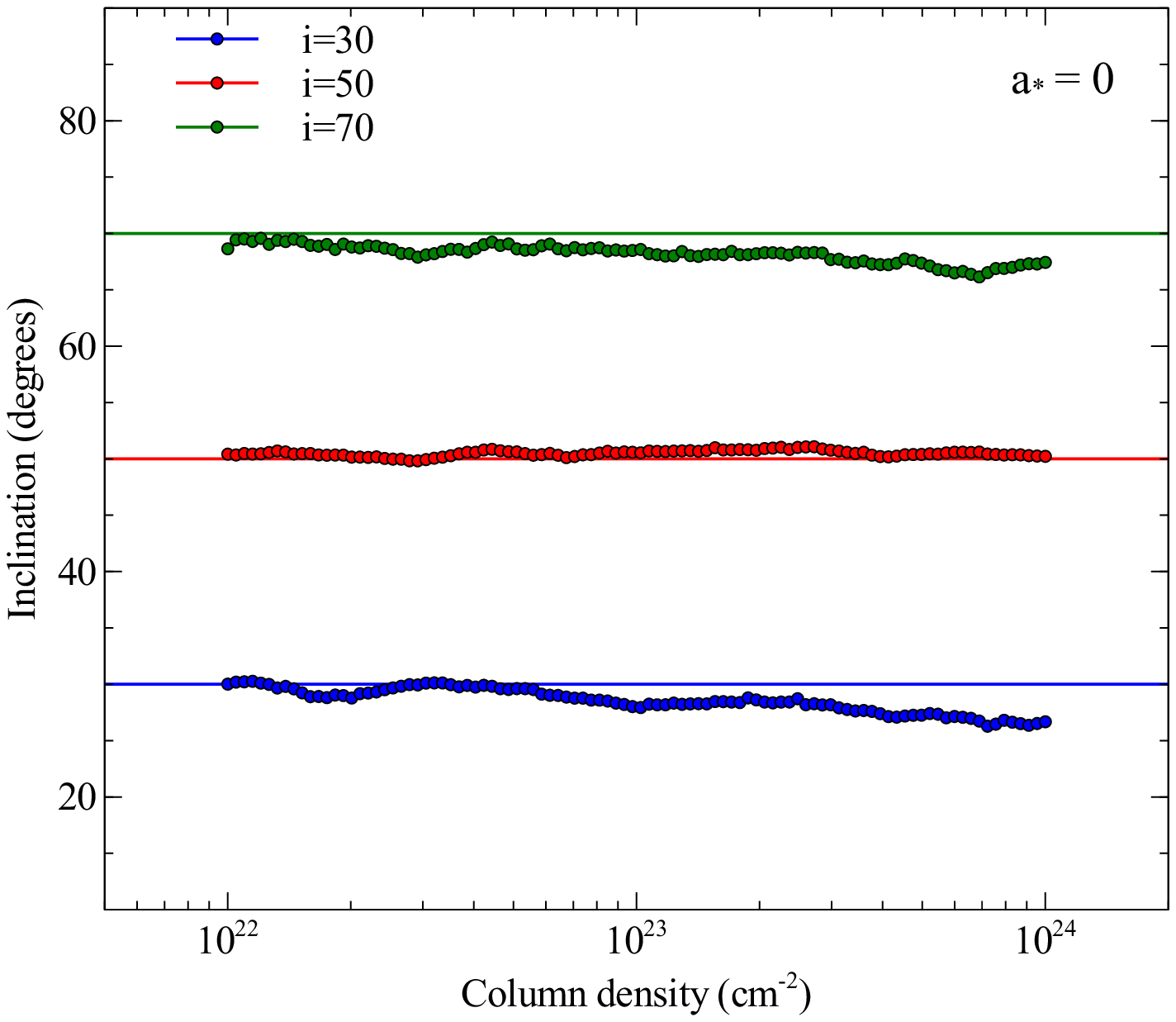}
\includegraphics[width=3in]{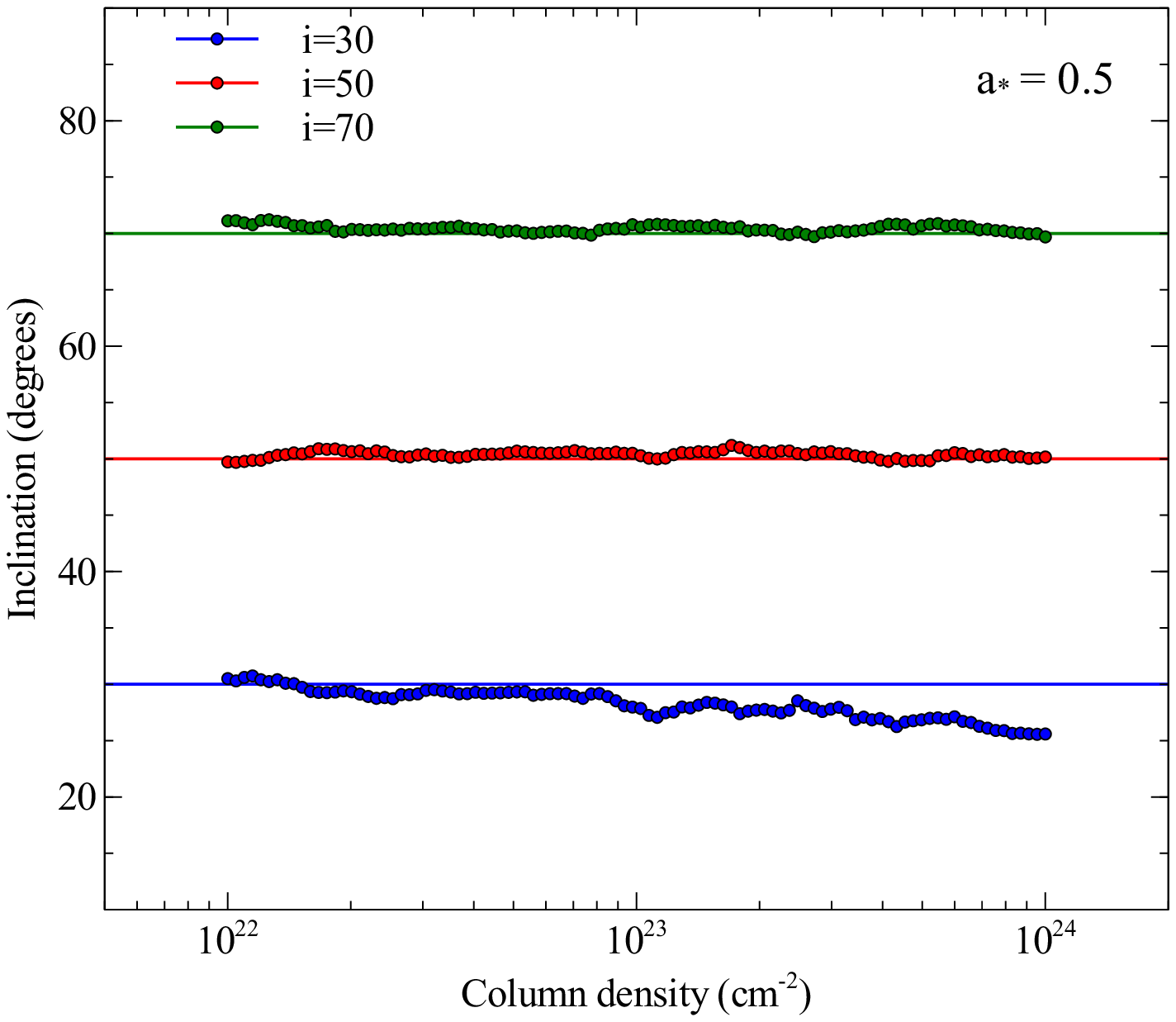}
\includegraphics[width=3in]{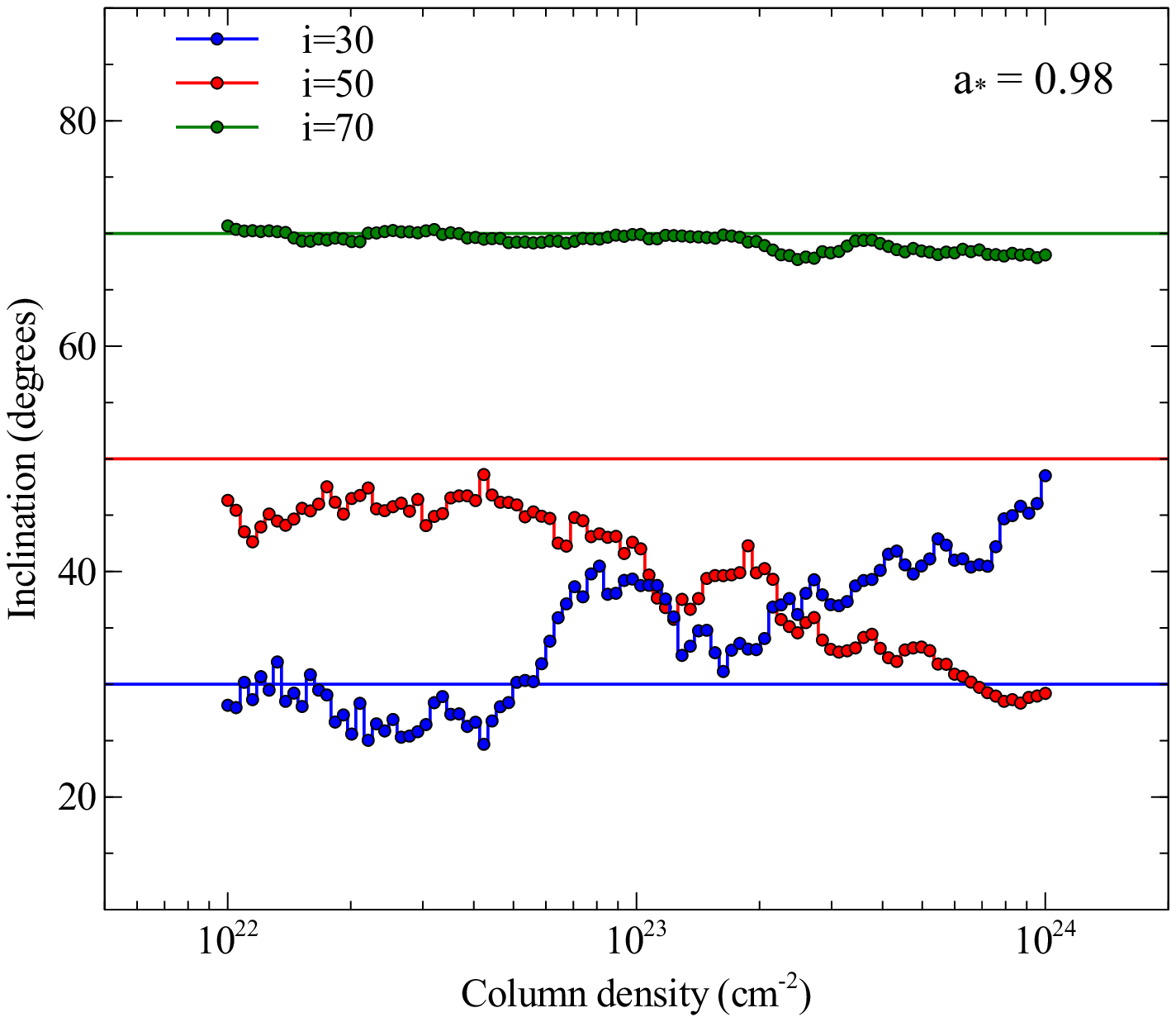}
\caption{The results of spectral simulations where we determine the distorting influence on the inner disc inclination that would result from {\it not} including the effect of absorption lines associated with Fe XXV/XXVI (with log$\xi$ = 3.5). The simulations are based on an input model from the {\it XMM-Newton/NuSTAR} study of NGC 1365 by Walton et al. (2014). We test for three inclinations at 30, 50 and 70\degree across a range in column density and for three black hole spins: a$_{*}$ = 0  (top-left), 0.5 (top-right) and 0.98 (bottom) with the results appropriately smoothed before plotting. As expected, the largest deviations occur for moderate inclinations at highest spin. 
}
\end{figure*}

\begin{figure*}
\includegraphics[width=3in]{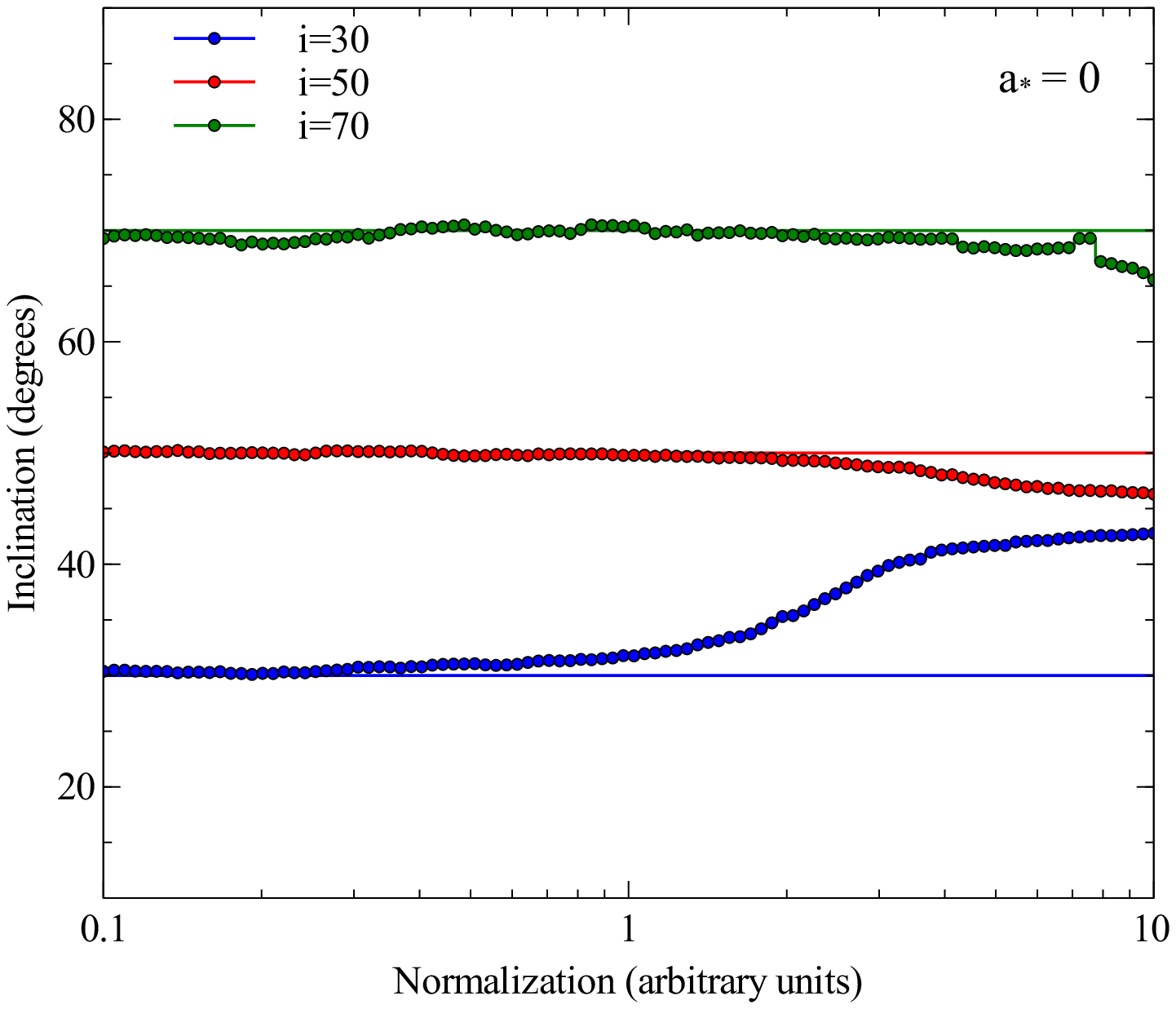}
\includegraphics[width=3in]{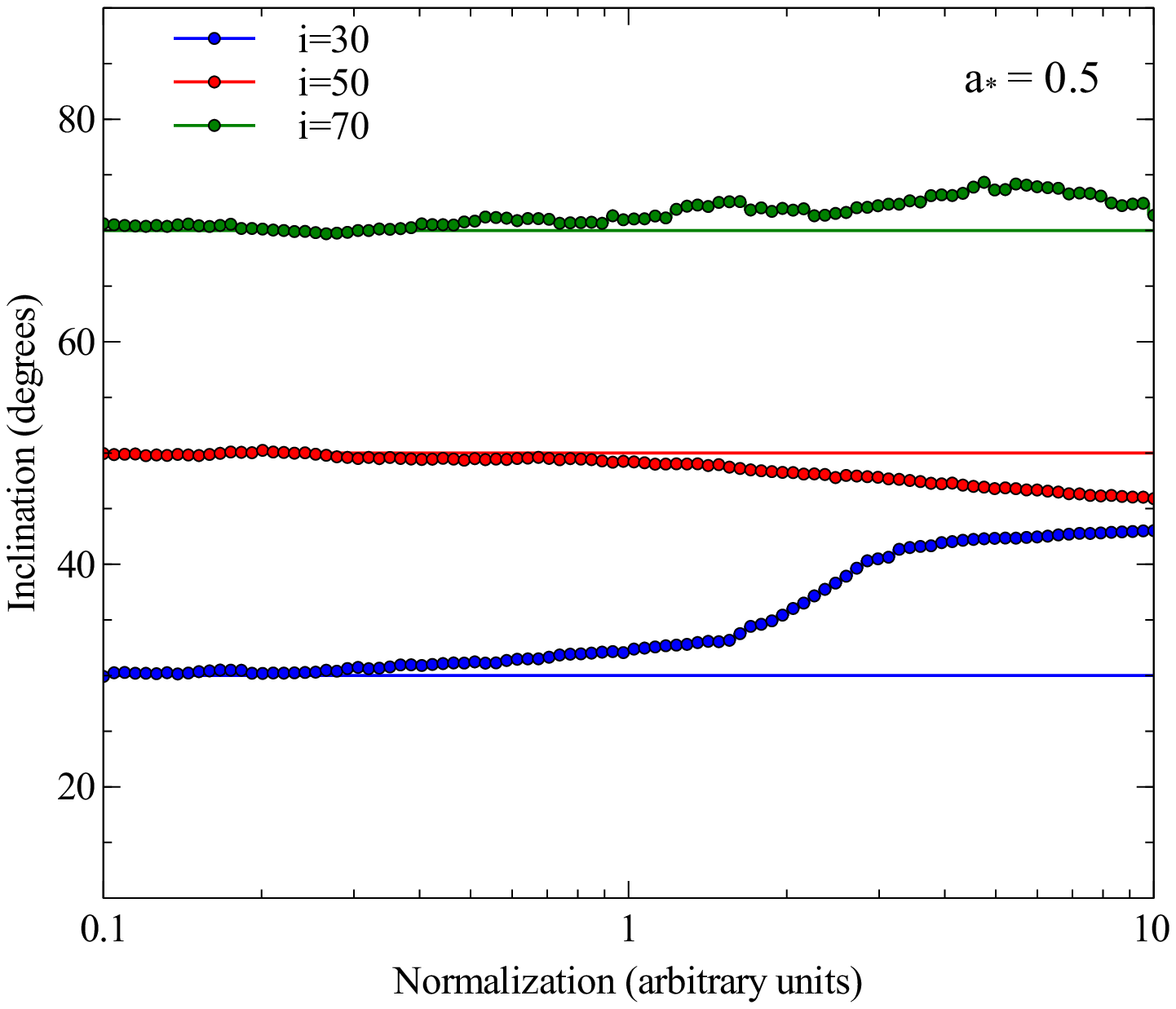}
\includegraphics[width=3in]{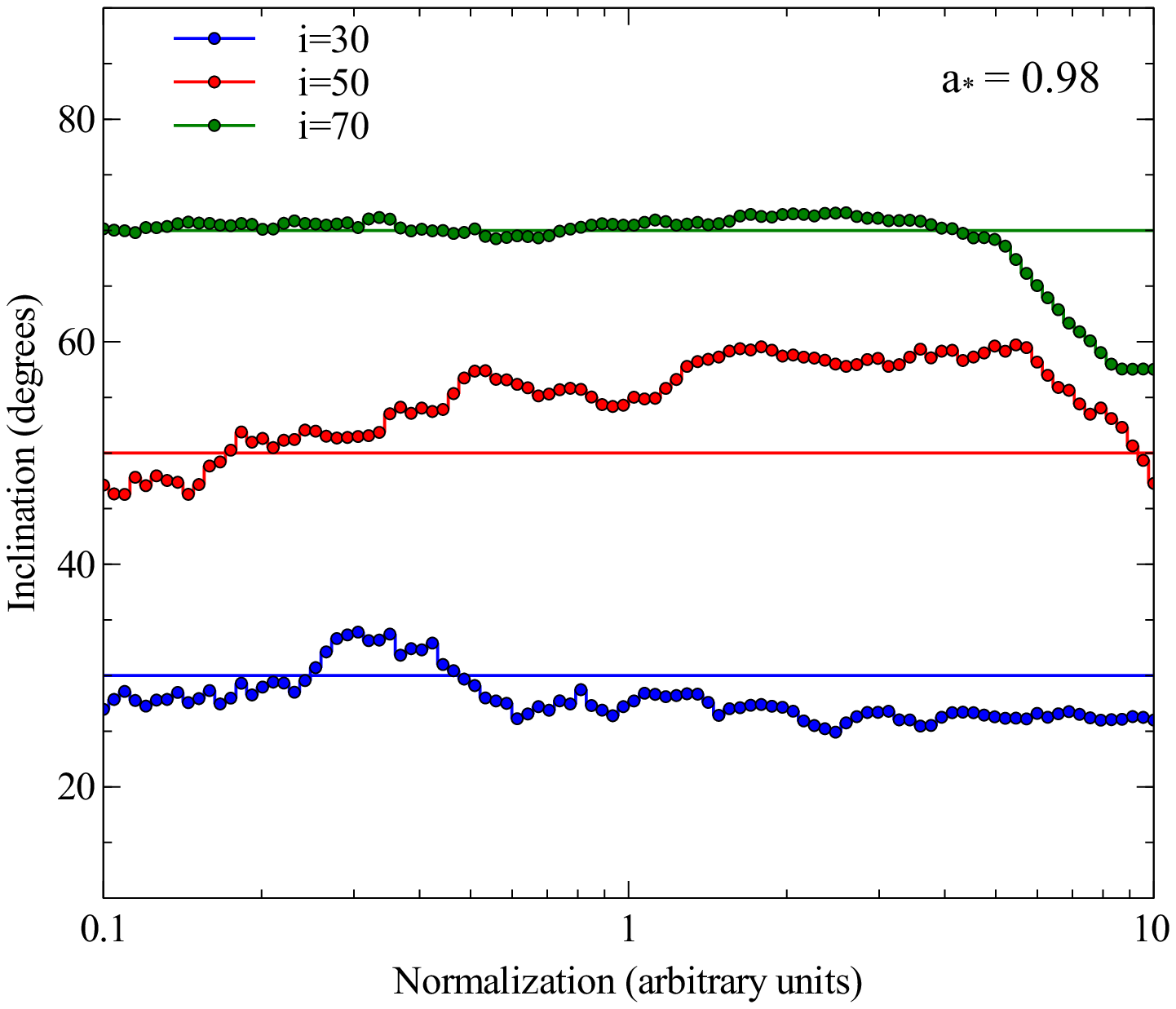}
\caption{The results of spectral simulations where we determine the distorting influence on the inner disc inclination that would result from {\it not} including the effect of emission lines associated with photo-ionised plasma (with log$\xi$ = 3.7: Lohfink et al. 2012). As before, the simulations are based on an input model from the {\it XMM-Newton/NuSTAR} study of NGC 1365 by Walton et al. (2014). We test for three inclinations at 30, 50 and 70\degree across a range in normalisation and for three black hole spins: a$_{*}$ = 0  (top-left), 0.5 (top-right) and 0.98 (bottom) with the results appropriately smoothed before plotting. 
}
\end{figure*}


\subsection{Uncertainty on the inner disc inclination} 

A potential weakness in our use of inclinations obtained via reflection model fitting is the effect of unmodeled, narrow absorption and emission Fe XXV/XXVI lines associated with photo-ionised plasma (potentially outflowing, e.g. Tombesi et al. 2013). These lines have the potential to distort the shape of the blue wing of the Fe K$_{\alpha}$ line and thereby affect the derived inclination (as well as the measured black hole spin: see the discussion of Lohfink et al. 2012). For most of the AGN in our sample, such lines have been detected and modelled accordingly, however, where such lines cannot be ruled out (either due to model degeneracy or where data quality prevents their detection), the possibility remains that such lines may play a distorting role in our measurement of the inclination. Those sources where absorption/emission has not been included in the modelling are indicated in Table 1 by a $\dagger$ symbol.

To better quantify the potential distorting effect on the inclination that absorption or emission lines might have, we create simulated spectra using input models for the cases of absorption and emission separately. In both cases we parameterise the continuum and reflection using the {\sc xspec} model {\sc relxill} (Dauser et al. 2013) with the input parameters (primary photon index, ionisation state etc.) fixed to those found in the {\it XMM-Newton} and {\it NuSTAR} study of NGC 1365 (Walton et al. 2014). We then introduce rest-frame, narrow absorption lines (using an {\sc xstar} grid - Walton et al. 2013), with the ionisation state set at that for NGC 1365 (log$\xi$ = 3.5: Risaliti et al. 2013) or for the case of emission, a series of photo-ionised emission lines (through inclusion of a bespoke {\sc xstar} grid made using {\sc xstar2xspec} based on the assumed illuminating SED and stepping through a range in ionisation state and column density). Whilst some AGN show blue-shifted absorption lines (notably PDS 456: Nardini et al. 2015), by choosing the absorption (and emission) lines to be rest-frame, we impose the largest amount of overlap between the narrow features and the blue wing of the Fe K$_{\alpha}$ line and so the most conservative estimate of the distortion to our inclination measurements.

We proceed to step our input models through inclinations of 30, 50 and 70\degree; at smaller inclinations the observed energy of the Fe K$_{\alpha}$ line becomes increasingly dominated by gravitational redshift rather than Doppler shifting (and so the effect of absorption/emission around the blue wing of the line is typically lessened for moderate to high spins), whilst at inclinations $>$ 70\degree we expect the molecular torus to dominate the line-of-sight. We simulate for a set of spin values (a$_{*}$ = 0, 0.5 and 0.98) which determines the amount of Doppler-shift (for a given inclination) and gravitational redshift in the Fe K$_{\alpha}$ line. As we are performing a lower resolution investigation compared to our comparison of reflection models, for the low and moderate spins we set the emissivity to be R$^{-3}$ and reflection fraction to be unity, and for the highest spin we set the emissivity to be R$^{-7}$ with a reflection fraction of 2 to account for the increased inner disc area and stronger light-bending (see Wilkins \& Fabian 2012). For the case of Fe XXV/XXVI absorption lines, we step through a range in column density which in turn provides a range in equivalent width (as we assume we are on the unsaturated, linear part of the curve of growth) and for emission we step through a range in normalisation (equivalent to a step in column density) for a fixed ionisation state (log$\xi$ = 3.7 and assuming unity covering fraction) based on a study of Fairall 9 where such emission lines have been well modelled (Lohfink et al. 2012).

For each input value of spin, inclination and column density/normalisation, we simulate data (using the {\sc fakeit} tool in {\sc xspec}) for the nominal {\it XMM-Newton}, 0.3-10~keV bandpass, and for a representative exposure time of 100~ks. After simulating each spectrum, we re-estimate the inclination by modelling the data {\it without} accounting for the narrow Fe XXV/XXVI lines in absorption or emission respectively; in doing so we obtain a crude indication of possible inaccuracies in our measured inclination values. As we are only interested in modelling the Fe K$_{\alpha}$ line, we fit across the 3-10~keV range (and so avoid any complicating effects of neutral absorption column and the soft excess that are important for modelling of real datasets). 

The results of our simulations are shown in Figures 3 \& 4 which demonstrate that the maximum distortion to measurements of the inner disc inclination as a result of unmodeled {\it absorption} features occurs at highest spins and moderate inclinations. This is a result of the broadened blue wing of the Fe K$_{\alpha}$ line profile which increasingly overlaps with the absorption lines and therefore has a larger impact on the inferred inclination as we move to higher spins. This distortion is somewhat mitigated at the highest inclinations due to the increased Doppler boosting and shifting of the blue wing. In the case of narrow {\it emission} features, the distortion at lowest inclinations is now greatest for low spin whilst for higher inclinations is greatest at higher spins. At low inclinations, the shift in the Fe K$_{\alpha}$ line is dominated by gravitational redshift and so the higher the spin the lower energy the line centroid; thus at low spins the centroid energy is higher and therefore closer to the narrow Fe XXV/XXVI emission lines, leading to possible inaccuracies in the inferred inclination. This is markedly different to the case of absorption where the lines are at higher energies and so the overlap (and therefore distortion) at low spin is decreased. At higher inclinations - as with the case of absorption - the role of the blue wing of the Fe K$_{\alpha}$ line becomes most important and at higher spins has greater overlap with the narrow emission lines leading to increased distortion of the true inclination.

\begin{figure*}
\includegraphics[width=5in]{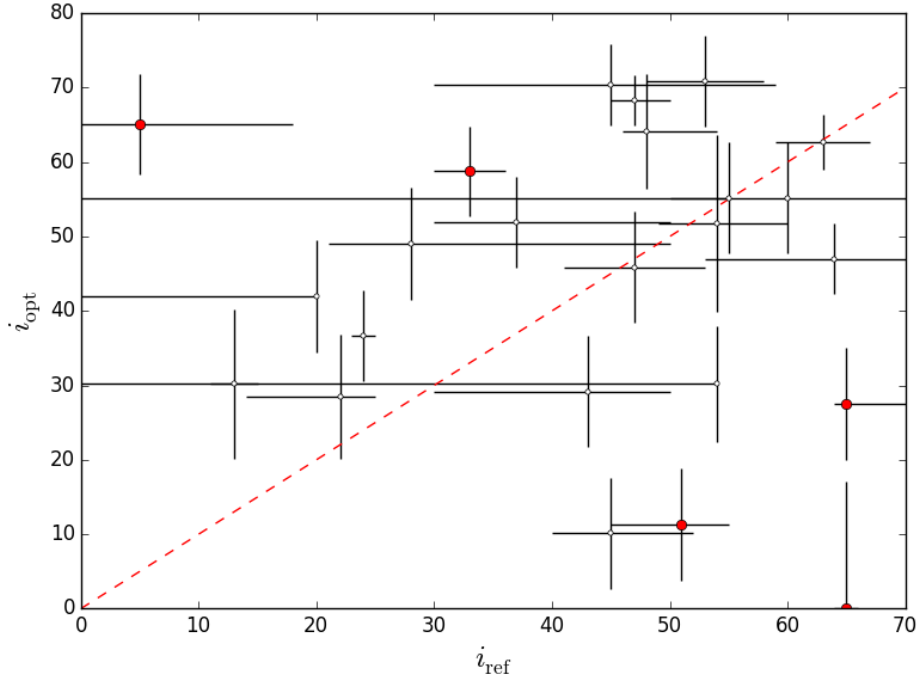}
\caption{The AGN inner disc inclination ($i_{\rm ref}$) derived from fitting the reflection spectrum plotted against the inclination of the host galaxy stellar disc ($i_{\rm opt}$). The red line indicates a 1:1 correlation and we find, from performing a Pearson product-moment correlation test against the entire sample and Monte-Carlo simulations, that the sample is significantly ($\gg$ 5-$\sigma$) uncorrelated. If we assume that the sample is {\bf not} homogeneous then we find  that by removal of the five highlighted sources (red points), we can reproduce (at 3$\sigma$) the remaining sample distribution from an underlying 1:1 correlation. }

\end{figure*}

Although the spin distribution of AGN (see Reynolds 2013) would appear to suggest that most SMBHs tend towards maximal prograde (see the discussions on selection effects in Brenneman et al. 2011 and Vasudevan et al. 2015), we make no limiting assumption and only focus on the maximum distortion possible to the derived inclination of those AGN where the emission/absorption has not been modelled (in many cases this is simply a result of the data not supporting the inclusion of such an extra component: Walton et al. 2013). From inspection of Figures  3 \& 4 we expand the error bounds for those sources in Table 1 indicated by a $\dagger$, to encompass the maximum, {\it approximate} range in value the true inclination might take based on our simulations; for example, in the case of IRAS 13224-3809, a measured inclination of 65\degree is compatible with an underlying inclination of 70\degree and so we set the positive error to be 5\degree rather than 1\degree.

There are a number of key assumptions upon which our simulations (and conservative error bounds) rest, namely the input model parameters and how representative of our sample of AGN these really are. As we have simulated absorption and emission features for a given column density (and therefore equivalent width), we can essentially ignore the effect of changing Fe abundance (which would only act to scale the strength of both the Fe K$_{\alpha}$ line and narrow lines respectively). Whilst it is not clear how representative our assumed ionisation states for the emitting and absorbing plasmas are, we have based these on the detailed studies of NGC 1365 (Walton et al. 2013) and Fairall 9 (Lohfink et al. 2012) which benefited from very high quality data in the bands of interest; further studies of such lines in a larger number of sources will no doubt allow us to revisit this aspect of the modelling.

Finally, whilst our choice of emissivity profiles is consistent with the illuminating source likely being associated with the base of a jet (e.g. Wilkins \& Gallo 2015), steeper indices have been reported (e.g. ESO 113-G010: Cackett et al. 2013); in such cases the blue wing of the iron line should be proportionally stronger (as more of the primary continuum is focused onto the most inner regions which are rotating fastest) which should make mis-modelling less likely. 
There are no doubt other potential issues which we cannot yet address including the effect of the unknown vertical and radial structure of the disc itself, however, this is still somewhat of an open question until the effect of magnetic pressure in the disc is fully understood (see e.g. Nayakshin, Kazanas \& Kallman 2000; Blaes, Hirose \& Krolik 2007; Fabian \& Ross 2010). 


\subsection{Host stellar disc inclination}

For the majority of sources, the inclination of the host stellar disc is taken from the {\sc HyperLEDA} database\footnotemark\footnotetext{http://leda.univ-lyon1.fr} (see Makarov et al. 2014) and is derived from the Hubble formula:
\begin{equation}
\sin^{2}(i) = \frac{1-10^{-2\log r_{25}}}{1-10^{-2\log r_{\rm o}}}
\end{equation}
where $\log r_{\rm 25}$ is the logarithm of the ratio of semi-major to semi-minor 25 mag/arcsec$^{2}$ isophotes in the  B-band, and $\log r_{\rm o}$ is a function of $t$, the morphological type (taking values from -5 to 10 for ellipticals through spirals and to irregulars: De Vaucouleurs 1959): $\log r_{\rm o}$  = 0.43 + 0.053$t$ for -5 $< t <$ 7 and 0.38 for $t >$ 7. Where inclinations are not available from the database, we search the literature for the sizes of the isophotes and morphological type and derive the values ourselves. In total we obtain inclinations for the host stellar disc in 26 out of the 29 AGN. Errors on the inclination are derived from the uncertainties on $logr_{\rm 25}$ and morphological type. Where the morphological type is unknown, the values in the {\sc HyperLEDA} database assume that the galaxy is an Scd - type 6 of De Vaucouleurs (1959) - as this is the most common morphology; in these cases we assume the mean error on the inclination obtained from those sources where uncertainties on the morphology are provided. We note the implicit assumption that the B$_{\rm 25}$ isophotes have been correctly modelled and account for any contamination by the host AGN - confirming this via individual modelling of optical profiles is beyond the scope of this paper. 

\subsection{Correlation analysis} 

We plot the inclinations of the inner disc from reflection fits against the stellar disc inclinations in Figure 5. We proceed to test the degree of correlation using the Pearson product-moment correlation coefficient, finding an $r$ value (which takes values between -1 for a perfect anti-correlation to +1 for a perfect correlation) of -0.1, indicating a lack of a correlation (with a p-value of 0.6). 
In order to determine the likelihood of obtaining this result by chance, we perform a large series of Monte-Carlo simulations to determine how often such a weak correlation would result from an underlying 1:1 correlation (i.e. the AGN inner disc and host galaxy being aligned) given the size of the errors on the respective inclinations. We find that, even from 10 million simulations, we should never expect such a weak correlation, indicating a lack of correlation across the {\it entire sample} at $\gg$ 5$\sigma$. 

Our correlation test above implicitly assumes that the population of AGN in our sample is homogeneous with respect to the feeding mechanism driving the recent phase of activity, however, there is no reason to believe that this should be the case. Indeed a large number of the sources in our sample would appear to sit within 3$\sigma$ of the 1:1 correlation in Figure 5. To isolate outliers from our sample we remove sources one-by-one which contribute to the low product-moment correlation coefficient (starting with the largest contributors), and, for each source removed from the sample, we repeat our correlation analysis until the remaining sample is consistent within 3$\sigma$ of originating from an underlying 1:1 correlation. We find that a total of five sources have to be removed (shown in Figure 5 as red points), these are (with their corresponding delta product-moment correlation coefficients): Mrk~335 (0.127), 3C~120 (0.108), IRAS~13224-3809 (0.044), 1H~0419-577 (0.026), 1H~0707-495 (0.022) with the next highest contributor being MCG-6-30-15 (0.019).

\section{Discussion \& Conclusion}

In Table 1 we present our compiled sample of 29 AGN with measurements of the inclination of the inner disc via recent reflection model fitting to the X-ray spectrum. We explore possible sources of error introduced through use of different models for the reflected emission ({\sc reflionx} or {\sc relxill}) finding that, on average, the expected error is smaller than the measurement errors. We also explore the distorting influence on the inclination resulting from unmodeled emission and absorption by ionised species of Fe and increase the error bounds accordingly. When combined with host galaxy inclinations, derived from the ratio of semi-major to semi-minor axis and host morphology, we obtain a sub-sample of 26 sources and a significant ($\gg$ 5$\sigma$) lack of a correlation between the inclinations. If we assume that the population within our sample is homogeneous with respect to the feeding mechanism, this would be consistent with the situation where minor-mergers were driving the activity; either the SMBH is aligned with the host galaxy and the infalling gas from the merger has formed an inner disc which is not yet aligned with the SMBH via the BP effect (see Natarajan \& Pringle 1998; King et al. 2005), or - more likely - the SMBH is misaligned with the host galaxy (as a result of past mergers) and the infalling material is aligned with the SMBH via the BP effect. Alternatively, the SMBH may be misaligned with the host galaxy and is being fed via the stellar disc but not enough angular momentum has been transferred to the black hole to align it, and so the BP effect produces a misaligned inner disc. Finally, and assuming homogeneity, the entire population could be fed via co-planar accretion but illumination by the corona produces a radiative disc warp (Pringle 1996, 1997).

There is no {\it a priori} expectation that the AGN activity we observe should result from material fed into the inner sub-pc regions via the {\bf same} mechanism in all sources. By selecting off those sources which contribute to the poor correlation we obtain a sample of five outliers from a sample of twenty-one which are consistent with an underlying 1:1 correlation (at the 3$\sigma$ level). This might indicate that there are at least two populations at work, most ($\sim$4/5) of which could be fed by co-planar accretion with little or no distortion via a radiative disc warp and a sub-sample which could be co-planar with a disc warp or fed (now or in the past) by minor mergers (possibly with a disc warp as well). If both sub-samples are indeed fed by coplanar accretion but we see radiative disc warps in only one then we should expect there to be some obvious differences in the illumination properties and therefore the emission spectrum, yet these do not appear to be substantially different (e.g. Walton et al. 2013). We therefore arrive at the conclusion that, should there be multiple underlying populations in our AGN sample, that we are probably seeing the effect of differing feeding mechanisms and the BP effect rather than radiative warping. As further supporting evidence, we note that in the case of MCG-6-30-15 (which is close to being in our sample of outliers), Raimundo et al. (2013) have discovered that the inner $\sim$100~pc is counter-rotating, which could imply a merger history (although see Algorry et al. 2014 for an alternative mechanism). By extension, our sample of five `outlier AGN', might indicate that misalignment is rather commonplace as suggested by other authors by other means including from the angle of the jet (e.g. Clarke, Kinney \& Pringle 1998; Nagar \& Wilson 1999; Kinney et al. 2000, although see also Walker, Bagchi \& Fabian 2015 for the discovery of FRII jets in a massive spiral galaxy which appear aligned with their host) and from the inclination of the narrow line region (e.g. Fischer et al. 2013). 

Should two populations truly be present, our analysis might imply that the majority of AGN activity results from co-planar accretion through the host stellar disc. We note that, whilst it is plausible that this could result from feeding a misaligned SMBH via mergers, the time-scales for alignment of the inner disc with the SMBH are expected to be short (Natarajan \& Pringle 1998) and so a systematic tendency towards apparent alignment seems unlikely. However, we caution that our selection criteria is naturally biased towards those AGN which have high quality data around the Fe K$_{\alpha}$ line; as this sits in an energy range where typical detector responses are falling away, we are likely to be systematically selecting those sources at the brighter end of the AGN population. As a result, there may be an unintended selection bias towards AGN with highly spinning SMBHs, as the radiative efficiency (and therefore luminosity for a given mass accretion rate) is substantially higher (e.g. Brenneman et al. 2011; Vasudevan et al. 2015). As the highest spins are predicted to result from long-lived, coherent accretion (e.g. Fanidakis et al. 2011, although see also Dotti et al. 2013), the observation that the majority of our sources may be fed via this process is not necessarily representative of the larger AGN population. However, any such conclusions about the feeding mechanism naturally makes the assumption that accretion via the stellar disc is indeed co-planar; if material is instead fed into the inner sub-pc regions from galactic winds (resulting from supernovae in starburst galaxies, e.g. Tenorio-Tagle, Silich, Munoz-Tunon 2003), then the infalling material's angular momentum need not be fully aligned with the host galaxy. However, such gas would be assumed to carry with it some component of the angular momentum of the stellar disc and should therefore lead to only increased scatter about a 1:1 correlation.

At present we cannot be certain that our sample is heterogenous although there is some supporting evidence of minor mergers feeding at least some AGN in the local Universe (e.g. MCG 6-3015: Raimundo et al. 2013; Mkn 509: Fischer et al. 2015). To investigate this fully requires detailed optical studies of statistical outliers and a larger sample size to increase their number; the introduction of next-generation X-ray missions such as {\it eROSITA} and {\it Athena} will dramatically expand the sample size of AGN with very high quality data for spectral modelling and population analyses.  A merger history leading to misaligned SMBHs and inner discs is also expected to leave an imprint in the orientation of kpc jets (see Kinney et al. 2000 for evidence of this) which are expected to be launched along the black hole spin axis (Blandford \& Znajek 1977 but see also Natarajan \& Pringle 1998). Such jets are expected to have been launched during each accretion epoch; over time we might therefore expect to see enriched material in a large solid angle or a number of synchrotron `cavities' in various orientations mapping out the accretion history; our sample of five `misaligned' AGN therefore represent excellent targets for more detailed multi-wavelength studies. 

Finally, we expect future theoretical and analytical developments to improve the constraints on the level of systematic alignment/misalignment in the larger AGN population. With respect to the future of spectral modelling and the measurement of the inner-disc inclination, the continuing development of radiative 3D-MHD codes (e.g. Jiang, Stone \& Davis 2014) will allow a much deeper physical understanding of the structure of the accretion disc and the role (important or otherwise) that magnetic pressure may play in hydrostatic balance and it's impact on the reflection spectrum. The increasing availability of data extending to high energies (thanks to {\it NuSTAR}) allows the contribution and shape of the Compton hump to be well modelled (e.g. Risaliti et al. 2013) and the parameters of the reflection models (e.g. the reflected fraction, ionisation state and inclination) to be better constrained in the wider population of AGN (e.g. Marinucci et al. 2014). Finally, the introduction of advanced techniques that fully utilise the time domain (e.g. modelling of the Fourier phase lags via impulse response functions: Cackett et al. 2014; Emmanoulopoulos et al. 2014; Uttley et al. 2014) are expected to significantly help constrain the geometry of the system and thereby allow for the most rigorous measurements of the inner-disc inclination.

\section{Acknowledgements}

We thank the anonymous referee for their useful comments and suggestions. We also thank James Miller-Jones for valuable insight into VLBI. MJM appreciates support from ERC grant 340442. MLP appreciates support from an STFC PhD grant. CSR is grateful for financial support from the Simons Foundation (through a Simons Fellowship in Theoretical Physics), a Sackler Fellowship (hosted by the Institute of Astronomy, Cambridge), and NASA under grant NNX14AF86G.  CSR thanks the Institute of Astronomy at the University of Cambridge for their hospitality during an extended visit while this work was being performed.

\label{lastpage}

\end{document}